\documentclass[12pt,a4paper]{article}
\RequirePackage{amsthm,amsmath,amsfonts,amssymb}
\RequirePackage[numbers]{natbib}
\RequirePackage[colorlinks,citecolor=blue,urlcolor=blue]{hyperref}
\RequirePackage{graphicx}
\usepackage{newfloat}
\usepackage{listings}
\usepackage{booktabs}
\usepackage{authblk}
\usepackage{subfig}
\graphicspath{{figures/}}
\theoremstyle{plain}

\newtheorem{theorem}{Theorem}[section]
\newtheorem{lemma}[theorem]{Lemma}
\theoremstyle{remark}
\newtheorem{remark}{Remark}

\newcommand{\ba}{\begin{array}}\newcommand{\ea}{\end{array}}
\newcommand{\be}{\begin{eqnarray}}\newcommand{\ee}{\end{eqnarray}}
\newcommand{\beq}{\begin{equation}}\newcommand{\eeq}{\end{equation}}
\newcommand{\bex}{\begin{eqnarray*}}
\newcommand{\eex}{\end{eqnarray*}}

\newcommand{\wh}{\widehat}
\newcommand{\op}{\operatorname}
\newcommand{\f}[2]{\displaystyle\frac{#1}{#2}}

\newcommand{\dr}{\stackrel{\mathcal{D}}{\longrightarrow}}
\newcommand{\ind}{\rotatebox[origin=c]{90}{$\models$}} 
\newcommand{\mb}{\mathbb}
\newcommand{\rr}{\rightarrow}
\newcommand{\la}{\langle}
\newcommand{\ra}{\rangle}

\allowdisplaybreaks

\begin{document}

\title{
An adaptive model checking test for functional linear model}
\author[1]{Enze Shi}
\author[1]{Yi Liu}
\author[1]{Ke Sun}
\author[1]{Lingzhu Li}
\author[1]{Linglong Kong}
\date{}
\affil[1]{Department of Mathematical and Statistical Sciences,
University of Alberta}

\maketitle

\begin{abstract}
Numerous studies have been devoted to the estimation and inference problems for functional linear models (FLM).
However, few works focus on model checking problem that ensures the reliability of results.
Limited tests in this area do not have tractable null distributions or asymptotic analysis under alternatives.
Also, the functional predictor is usually assumed to be fully observed, which is impractical.
To address these problems, we propose an adaptive model checking test for FLM.
It combines regular moment-based and conditional moment-based tests, and achieves model adaptivity via the dimension of a residual-based subspace.
The advantages of our test are manifold.
First, it has a tractable chi-squared null distribution and higher powers under the alternatives than its components.
Second, asymptotic properties under different underlying models are developed, including the unvisited local alternatives.
Third, the test statistic is constructed upon finite grid points, which incorporates the discrete nature of collected data.
We develop the desirable relationship between sample size and number of grid points to maintain the asymptotic properties.
Besides, we provide a data-driven approach to estimate the dimension leading to model adaptivity, which is promising in sufficient dimension reduction.
We conduct comprehensive numerical experiments to demonstrate the advantages the test inherits from its two simple components.

\end{abstract}





\section{Introduction}

As an important component of functional data analysis (FDA), FLM is widely adopted in practice to describe the relationship between a functional predictor and a scalar response. It has been actively studied and received increasing attention in recent decades \cite{ramsay2005functional,ferraty2006nonparametric,yuan2010reproducing}. Classical FLM can be formulated as
\be
\label{flm}
Y = \int_{\mb{I}}X(t)\beta(t)dt+\eta,
\ee
where $Y\in\mb{R}$ is a scalar response, $X(\cdot) \in L^2(\mb{I})$ is a real-valued random process over the interval $\mathbb{I}=[a,b]$, $\beta(\cdot)$ is an unknown slope function in $L^2(\mb{I})$, and $\eta$ is a random noise satisfying $\mb{E}(\eta\mid X(\cdot) )=0$. Without loss of generality, let $\mb{I}=[0,1]$ and assume $Y$ and $X(\cdot)$ are centered.
Apart from the scalar-on-function model in \eqref{flm}, other forms of FLM include function-on-function regression \cite{yao2005functional,sun2018optimal} and function-on-scalar regression \cite{reiss2010fast,goldsmith2016assessing,zhang2021high}, and also generalized FLM \cite{muller2005generalized,shang2015nonparametric}.

There are extensive investigations into estimation \cite{cai2006prediction,cai2012minimax,cai2011optimal,hall2007methodology} and inference \cite{shang2015nonparametric,lei2014adaptive,cardot2003testing,xue2021hypothesis} problems for FLM. However, most of them assume the model is sufficient. A wrongly specified model could lead to unreliable conclusions, making the model checking procedure an essential step before fitting the data. Even though much attention has been paid to this area, few theoretical results on model checking problems for functional data are developed. Limited tests for FLM include scalar-on-function regression \cite{garcia2014goodness} and function-on-function regression \cite{chen2020model}, both of which are motivated by the residual-marked empirical process proposed in \cite{escanciano2006consistent}. A recent work \cite{cuesta2019goodness} considers an efficient empirical process-based test using random projection for scalar-on-function linear regression. It greatly reduces the complexity of calculating the test statistics but at the expense of lower power compared to \cite{garcia2014goodness}.
Nevertheless, resampling techniques such as wild bootstrap are still required to determine the critical value, which is a computation burden.
In addition, the discrete nature of collected data is usually disregarded, which can affect the convergence rate of the test statistic and impair its power.
Existing tests considering discretely observed data either lack the theoretical results under local alternatives or cannot provide a reference relationship between the sample size and the number of grid points for the asymptotic properties \cite{chen2020model,cai2011optimal,zhu2012multivariate}.
The demand for reducing the computation complexity, admitting the discretely observed data and establishing comprehensive theories drives our work.

We propose an adaptive model checking test for FLM and illuminate its asymptotic properties in different underlying models to address the challenges.
Our test is motivated by the adaptive-to-model hybrid test proposed in \cite{li2021adaptive}. We are interested in the incomplete nature of collected data. Assume the functional predictor $X(t)$ is observed at $M$ grid points on its support. We then study a desired $M$ to maintain the asymptotic properties rather than assuming $X(t)$ is completely observed as most existing works, which is more practical and realistic. For notation simplicity, denote $X = X(t)$, $\beta = \beta(t)$ and $\int_0^1 X(t)\beta(t)dt$ as $\la X,\beta\ra$ without confusion. Our objective is to test the following hypothesis
\be
\label{H0}
H_0:\quad Y = \la X,\beta_0\ra+\eta,\quad\text{for some }\beta_0\in L^2(\mb{I}),
\ee
against
\be
H_1:\quad  Y= G(X)+\eta,
\ee
where $E(\eta \mid X) = 0$ and $G(X)\neq \la X,\beta\ra$ for all $\beta\in L^2(\mb{I})$. Our test contains two major components. The first component simply uses a moment-based sum of weighted residuals as a test statistic. It shares asymptotic behaviors with global smoothing methods \cite{stute1998bootstrap,dette2007new} that can achieve the fastest possible convergence rate \cite{guo2016model,li2021adaptive} while having a tractable null distribution according to the inference results in \cite{shang2015nonparametric}.
The second component adjusts the typical conditional moment-based test proposed by \cite{zheng1996consistent} for functional data. It is sensitive to oscillating alternatives and can handling an omnibus test like other local smoothing methods \cite{hardle1993comparing,guo2016model}.
We can use an indicative dimension induced from a residual-based central mean subspace \citep{cook2002dimension,zhu2010sufficient,li2018sufficient} as a bridge to achieve model adaptivity and combine the merits of the two components.
The indicative dimension borrows ideas from sufficient dimension reduction (SDR) theory \cite{cook2002dimension,zhu2010sufficient,li2018sufficient}.
It has been applied to building adaptive-to-model tests \cite{guo2016model,tan2018projection,li2021adaptive} which can alleviate the curse of dimensionality. In the past decades, many efforts have been devoted to functional SDR \cite{ferre2003functional,ferre2005smoothed,hall2006properties,ferraty2013functional,lian2015functional,song2019sufficient,song2021nonlinear}, paving the way to build the hybrid tests for FLM.

Our main contributions are multifold.
\begin{enumerate}
    \item
    The hybrid test has a chi-squared null distribution. Therefore, we do not need a resampling method to obtain the critical value. It reduces the computational burden for functional data analysis.
Our test has a fast convergence rate and omnibus property against the alternatives simultaneously, achieved by an adaptive-to-model dimension.
    \item We derive the minimum number of grid points to preserve the asymptotic properties of hybrid test by incorporating the discretely observed functional data.
    \item We systematically illuminate the asymptotic properties of the hybrid test under the null hypothesis, global alternatives, and local alternatives.
    \item We also develop a promising data-driven method for estimating the indicative dimension in practice. The results from various numerical studies show this method is robust to different data generating processes and the underlying models.
\end{enumerate}


The rest of this paper is organized as follows. In Section \ref{test_construction}, we propose the hybrid test statistic for FLM and introduce the estimation procedure for slope function utilizing eigen-system in Hilbert space.
The estimation of indicative dimension by SDR in functional space is illustrated as well.
Section \ref{asymptotic} elaborates the asymptotic properties of the test statistic under different hypotheses.
In Section \ref{numerical}, we present the finite sample powerfulness of the proposed test by various experiments and real data sets.

\section{Methodology}
\label{test_construction}

Let $\{X_i, Y_i\}_{i=1}^n$ be a sequence of independent and identically distributed (i.i.d.) random copies of $\{X, Y\}$. Contrary to many other literatures, we assume the functional predictor $X(t)$ is only observed at $M$ grid points $0=t_1<t_2<\ldots<t_M=1$ satisfying $\max_{1\leq j\leq M-1}\{t_{j+1}-t_j\}\leq C_0M^{-1}$ for some constant $C_0$. For simplicity, we consider the equal distance observations on $X=X(t)$, that is $t_i=(i-1)/(M-1),\,i=1,2,\ldots,M$. Consider the following hybrid test
\be\label{test statistic general}
T_n =\gamma_{n,M}\left| V_{0}^{2}I(\hat{q}=0)+V_{1} I(\hat{q}>0)\right|,
\ee
where $V_0$ and $V_1$ are two simple tests, $\gamma_{n,M}$ is the standardizing factor, and $\hat{q}$ is an estimated indicative dimension which we will elaborate later.

In the following, we elaborate on how to estimate the slope function and indicative dimension.

For the test statistics in \eqref{test statistic general}, we suggest the following explicit form.
Suppose $\beta^*={\arg \min }_{\beta \in \mathcal{H}}\{E(Y-\la X,\beta\ra)\}$, which can be consistently estimated by $\hat \beta$.
Let $\epsilon_i=Y-\la X_i, \beta^\ast\ra$ and $\hat{\epsilon}_i=Y_i-\la \widehat{X}_i,\hat{\beta}\ra$ be the estimation.
First, we consider $V_0 = \sum_{i=1}^{n} \hat{\epsilon}_{i} w(\widehat{X}_{i}) / n$
where $\hat X_i$ is a consistent estimator of $X_i$ and $w(\cdot)$ is a non-constant weight function . Then we use the conditional moment-based test $V_1=\sum_{i=1}^{n} \sum_{j \neq i, j=1}^{n} \hat{\epsilon}_{i} \hat{\epsilon}_{j}K_{h}(\la\widehat{X}_i-\widehat{X}_j,\widehat{\beta}) /(n(n-1))$, where $K_h(\cdot)=K(\cdot/h)/h$, $K(\cdot)$ is a one-dimensional kernel function and $h$ is the bandwidth.

Next, we introduce the details of the estimations needed for the test.
\subsection{Two-stage slope function estimation}
There are two strategies to estimate $\beta^*$ based on discretely observed functional data. One is to directly use the discrete samples $X_i(t)$ for estimation, see \cite{cai2011optimal,zhu2012multivariate,chen2020model,zhang2021high}, but it can be complicated to obtain inference results. The other is the two-stage estimation procedure \cite{hall2006properties,cai2011optimal}, which will be adopted in this paper. First, non-parametric methods are used to smooth the observations on each curve. Then the closed form of the reproducing kernel-based regularized estimator $\hat{\beta}$ can be derived by smoothed curves. Specifically, in the first stage, we apply the spline smoothing method to $\{X_i(t_j)\}_{j=1}^M$ in an $r$-th order Sobolev-Hilbert space to obtain
$$
\widehat{X}_i=\underset{g \in H^{r}[0,1]}{\arg \min }\left\{\frac{1}{M}\sum_{j=1}^{M}\left(X_{i}(t_j)-g(t_{j})\right)^{2}+\lambda_1\int_0^1\left[g^{(r)}(t)\right]^{2} d t\right\},
$$
where $\lambda_1$ is the smoothness parameter, $H^r$ is the $r$-order Sobolev space defined by
\be
\label{sobolev}
\begin{array}{l}
H^{r}[0,1]= \left\{f: [0,1] \mapsto \mathbb{R} \mid f^{(j)}, j=0, \ldots, r-1 \text{ are absolutely continuous}\right.\\[0.5em]
\left.\text{ and } f^{(r)} \in L^{2}[0,1]\right\},
\end{array}
\ee
where $f^{(j)}$ denotes the $j$-th derivative of $f$. The smoothed curves $\{\wh{X}_i(t)\}_{i=1}^n$ will be used for reproducing kernel-based estimation in the second stage.

Assume that the slope function $\beta^\ast$ resides in a Reproducing Kernel Hilbert Space (RKHS) $\mathcal{H}=H^{m}[0,1]$, the $m$-order Sobolev space defined by \eqref{sobolev} equipped with the norm $\|\cdot\|_{\mathcal{H}}$ given by $\|\beta\|_{\mathcal{H}}=\sum_{i=0}^{m-1}(\int_0^1\beta^{(i)}(t)dt)^2+\int_0^1(\beta^{(m)}(t))^2dt$.
We have the following regularized $\hat \beta$:
\be
\label{beta_n}
\wh{\beta}_{n,M,\lambda}=\underset{\beta \in \mathcal{H}}{\arg \min }\left\{\frac{1}{n} \sum_{i=1}^{n}\left[Y_{i}-\la \wh{X}_i,\beta\ra \right]^{2}+\f{\lambda}{2} J(\beta,\beta)\right\},
\ee
where $J(\beta, \widetilde{\beta})=\int_{0}^{1} \beta^{(m)}(t) \widetilde{\beta}^{(m)}(t) d t$ is a roughness penalty. The solution of \eqref{beta_n} has closed form, see \cite{wahba1990spline,yuan2010reproducing} for detailed derivation.

To make valid statistical inference for $\wh{\beta}_{n,M,\lambda}$, the eigen-system of $\mathcal{H}$ needs to be established. Both \cite{yuan2010reproducing} and \cite{shang2015nonparametric} describe the construction procedure. Here we briefly illustrate it.  Let the null space induced by the semi-norm $J(\beta, \widetilde{\beta})$ on $\mathcal{H}$ be $\mathcal{H}_{0}:=\{\beta \in \mathcal{H}: J(\beta,\beta)=0\}$,
which is a finite-dimensional linear subspace of $\mathcal{H}$. Denote by $\mathcal{H}_{1}$ its orthogonal complement in $\mathcal{H}$ such that $\mathcal{H}=\mathcal{H}_{0} \oplus \mathcal{H}_{1}$, where $\mathcal{H}_{1}$ also forms a RKHS.  Let $K(s, t)$ be the corresponding reproducing kernel of $\mathcal{H}_1$, and $C(s,t)$ be the covariance function of random variable $X(t)$. Then we apply spectral decomposition on both $K$ and $C$ such that
$$
K(s,t)=\sum\limits_{\nu=1}^{\infty}\rho_{\nu}\psi_{\nu}(s)\psi_{\nu}(t),\quad C(s,t)=\sum\limits_{\nu=1}^{\infty}\mu_{\nu}\phi_{\nu}(s)\phi_{\nu}(t),
$$
where $\rho_1\geq\rho_2\geq\ldots$ are the eigenvalues of $K(s,t)$ and $\psi_{\nu}$ the associated eigenfunctions, $\mu_1\geq\mu_2\geq\ldots$ the eigenvalues of $C(s,t)$ and $\phi_{\nu}$ the associated eigenfunctions. Define the new norm on $\mathcal{H}$ by $\|\beta\|_{\widetilde{K}}^2 = \la C\beta,\beta \ra+J(\beta,\beta)$, where $\widetilde{K} = (C+K^{-1})^{-1}$ is the reproducing kernel. Let $\widetilde{K}^{1/2}$ be the square-root kernel of $\widetilde{K}$ and $\Omega(s, t)=(\widetilde{K}^{1 / 2} C \widetilde{K}^{1 / 2})(s, t)$ be the product kernel. Conduct the spectral decomposition gives $\Omega(s, t)=\sum_{\nu=1}^{\infty} \widetilde{\rho}_{\nu} \widetilde{\psi}_{\nu}(s) \widetilde{\psi}_{\nu}(t)$. Let $\varphi_{\nu}^*=\widetilde{\rho}_{\nu}^{-1 / 2} \widetilde{K}^{1 / 2} \widetilde{\psi}_{\nu}$ and $\rho_{\nu}^*=\widetilde{\rho}_{\nu}^{-1}-1$, then we obtain the eigen-system $(\rho_{\nu}^*,\varphi_{\nu}^*)_{\nu=1}^{\infty}$.

\begin{remark}
\label{rem2}
Here we list some properties of eigenvalues shown above. Suppose $C(s,t)$ satisfies Sacks-Ylvisaker conditions of order $s$, see \cite{ritter1995multivariate}, then $\mu_{\nu}\asymp {\nu}^{-2(s+1)}$. Recall that $m$ is the order of sobolev space in the estimation procedure and thus $\rho_{\nu}\asymp {\nu}^{-2m}$. Notice that the eigen-system we construct satisfies Assumption A3 in \cite{shang2015nonparametric}, which implies $\rho_{\nu}^*\asymp \nu^{2k}$ and $k=m+s+1$. These orders will determine the convergence rate of $\hat{\beta}_{n,M,\lambda}$ and the standardizing factor $\gamma_{n,M}$ in test statistics.
\end{remark}

\subsection{Indicative dimension}

In this subsection, we construct the indicative dimension that integrates the components in hybrid tests. First, we introduce some basic notations and definitions about SDR for scalar-on-function model, see \cite{song2019sufficient} for more details. Consider the random variables $X\in L^2[0,1]$ and $Y\in\mb{R}$. If there exists a functional vector $B=(\theta_1(t),\ldots,\theta_q(t))^T \in \mathcal{H}^q$, such that $Y\ind X\mid\la B,X\ra$, where $\la B,X\ra = (\la \theta_1,X\ra,\ldots,\la\theta_q,X\ra)^T $, then the space $\operatorname{Span}\{B\}$ is called a sufficient dimension reduction subspace of $Y$ with respect to $X$. The intersection of all the dimension reduction subspaces is called the central subspace and denoted as $\mathcal{S}_{Y\mid X}$. The dimension of the central subspace is denoted as $\op{dim}(\mathcal{S}_{Y\mid X})$. If $\op{Span}\{B\}$ is the central subspace, then $\op{dim}(\mathcal{S}_{Y\mid X}) =\op{dim}( \op{Span}\{B\}) = q$. The definition mentioned above is a generalization of SDR for finite-dimensional $X\in\mb{R}^p$ \cite{cook2002dimension}. When the conditional independence is replaced by $Y\ind E(Y\mid X)\mid \la B,X\ra$, the corresponding subspace $\mathcal{S}_{E(Y\mid X)}$ is called the central mean subspace with dimension $\op{dim}(\mathcal{S}_{E(Y\mid X)})$. We consider the central mean subspace in this work.

Recall that $\epsilon_i=Y_i-\la X_i, \beta^\ast\ra$ and $\hat{\epsilon}_i=Y_i-\la \widehat{X}_i,\widehat{\beta}_{n,M,\lambda}\ra$ is its estimation.
Under the null hypothesis, $\epsilon=\eta$ and then $\operatorname{dim}(\mathcal{S}_{E(\epsilon \mid X)})=0$. Under the alternatives, the remainder $\epsilon=G(X)+\eta-\la X, \beta_0\ra=\Delta(X, \eta)$ and  $\operatorname{dim}(\mathcal{S}_{E(\epsilon \mid X)})>0$ since $E\{\Delta(X, \eta) \mid X\}$ is a nonconstant function of $X$. Let $q=\operatorname{dim}(\mathcal{S}_{E(\epsilon \mid X)})$ and $\hat{q}$ be its estimation.  When $\widehat{\beta}_{n,M,\lambda}$ is a consistent estimator, $\hat{q}$ should also be consistent. Under the null hypothesis, $\hat{q}$ equals to $0$ with a probability going to 1; under the alternatives, it converges in probability to a positive $q$. This expected property will perform as a bridge to combine two simple tests together and get a more powerful hybrid test. To ensure the properties mentioned hold, we require some basic assumptions:

\textbf{A1:} $E(Y^2)<\infty$ and $E(\|X\|_{L^2}^4)<\infty$.

\textbf{A2:} The covariance function $C(s, t)$ of $X$ is continuous on $\mathbb{I} \times \mathbb{I}$. Furthermore, for any $\beta \in$ $L^{2}(\mathbb{I})$ satisfying $C \beta=0$, we have $\beta=0$.

\textbf{A3:} There is a bounded linear operator $P_{B}(C)$: $\mathcal{H}\mapsto\mathcal{H}$ such that the linearity condition $E(X\mid\la B,X\ra)=P_{B}(C)X$ is satisfied.

\textbf{A4:} $\op{Var}(X\mid\la B,X\ra)$ is a constant operator on $\mathcal{H}$.

A1 is commonly required for the consistency and asymptotic properties of $\widehat{\beta}_{n,M,\lambda}$. A2 regularity condition guarantees that $\|\cdot\|_{\tilde{K}}$ is well defined. It also implies that the dimension of a subspace in $\mathcal{H}$ will be preserved after being applied by $C$. A3 and A4 are usually known as the linearity condition and constant variance condition under the SDR framework, see \cite{ferre2003functional,ferre2005smoothed,song2019sufficient} for more information and see \cite{li2021adaptive} for finite-dimensional case. With these assumptions, We imitate the convex combined matrix proposed in \cite{li2021adaptive} and develop the indicative operator on $\mathcal{H}$ as
$$
M^{c c}=E(\epsilon X) \otimes E\left(\epsilon X\right)+H H ,
$$
where $ H=E\left(\epsilon X\otimes X\right)$ and $X(t)\otimes Y(t)=X(s)Y(t)$. The indicative operator can be regarded as a bivariate function defined on $[0,1]\times[0,1]$ and $M^{cc}(s,t)\equiv0$ under the null hypothesis. Also, $M^{cc}(s,t)$ is continuous, symmetric, and square integrable, thus admits the spectral decomposition $M^{cc}(s,t)=\sum_{\nu=1}^{\infty}\lambda_{\nu}e_{\nu}(s)e_{\nu}(t)$, where $\lambda_{\nu}$ is its eigenvalue and $e_{\nu}$ is the associated eigenfunction. The lemma below guarantees that $\op{dim}(\mathcal{S}_{E(\epsilon\mid X)})$ can be obtained from $M^{cc}$.

\begin{lemma}
\label{lem_Mcc}
Let Assumptions A1 through A4 be satisfied, the indicative operator $M^{cc}$ satisfies $\op{Range}(M^{c c}) \subseteq\op{Range}(C \mathcal{S}_{E(\epsilon \mid X)})$, where $\op{Range}(\Gamma)=\{\Gamma\beta: \forall \beta\in\mathcal{H}\}$. If the number of non-zero eigenvalues of $M^{c c}$ is $q$, then we have $\op{Range}(M^{c c})=\op{Range}(C \mathcal{S}_{E(\epsilon \mid X)})$, indicating $\operatorname{dim}(\mathcal{S}_{E(\epsilon \mid X)})=q$.
\end{lemma}

The indicative operator is estimated by its sample analogue $\widehat{M}^{c c}=\hat{E}(\epsilon X) \otimes \hat{E}(\epsilon X)+\hat{H} \hat{H}$, where $\hat{E}(\epsilon X)= (\sum_{j=1}^{n} \hat{\epsilon}_{j} \widehat{X}_{j})/n$ and $\hat{H}=(\sum_{j=1}^{n} \hat{\epsilon}_{j} \widehat{X}_{j}\otimes \widehat{X}_{j})/n$. The consistency of $\widehat{\beta}_{n,M,\lambda}$ induces the consistency of the estimated indicative operator, which will be discussed in the next section. As the lemma suggests, the key objective is to determine or estimate the number of non-zero eigenvalues of $M^{cc}$. The criterion we use is a slight modification of the thresholding double ridge ratio (TDRR) method developed by \cite{zhu2020dimensionality}.

\begin{remark}
According to the above estimation, the standardizing factor $\gamma_{n,M}=n/\sigma_n^2$, where $\sigma_n^2=\sum_{\nu=1}^{\infty}w_{\nu}^{2} /(1+\lambda \rho^*_{\nu})^{2}$ and $w_{\nu}=\int_{0}^{1} \widehat{X}_w(t) \varphi_{\nu}^*(t) d t$. Here $(\rho^*_{\nu},\varphi_{\nu}^*)$ is the eigen-system established on $\mathcal{H}$ and $\widehat{X}_w=(\sum_{i=1}^{n} \widehat{X}_i w(\widehat{X}_i))/n$.
\end{remark}

\begin{remark}
With the expected asymptotic properties of $\hat q$, $T_n$ reduces to $\gamma_{n,M}V_0^2$ under the null hypothesis, which follows the chi-square distribution $\chi^2_1$.
While under the alternatives, $T_n$ jumps to $\gamma_{n,M}|V_1|$.
It inherits advantages from the two components. First, it has a tractable null distribution and diverges to infinity faster than using $V_1$ only. Second, the proposed test can detect local alternatives that converge to the null at a rate slower than $\gamma_{n,M}^{1/2}$.
Theorefore, it is more powerful than simply using $V_0$ or $V_1$.
This is the unique advantage of the adaptive-to-model hybrid test.
We will elaborate more details about the asymptotic properties of $T_n$ in the next section.
\end{remark}

\section{Asymptotic Properties}
\label{asymptotic}

We now investigate the consistency of $\hat{\beta}_{n,M,\lambda}$ and $\hat{q}$ and provide the asymptotic behaviors of $T_n$ under null and alternatives.
Listed below are the assumptions for the theorems.


\textbf{A5:} There exist constants $c_0 \in(0,1)$ and $M_0>0$ such that $E(e^{c_0\|X\|_{L^{2}}})<\infty$ and for any $\beta \in$ $\mathcal{H}$, $E(|\la X,\beta \ra|^{4}) \leq M_{0}[E(|\la X,\beta \ra|^{2})]^{2}$

\textbf{A6:} Assume $|w_{\nu}|\asymp 1$, define $M_a=\sum_{\nu=1}^{\infty}({w_{\nu}^2})/{(1+\lambda \rho^*_{\nu})^a}$ for $a=1,2$, then $M_1\asymp M_2$.

\textbf{A7:} The bandwidth of the kernel $h$ satisfies $h\rr0$, and also $n^{k/(2k+1)}h\rr\infty$, $n^{1/(2k+1)}h\rr0$ as $n\rr\infty$.

A5 is the regularity condition on process $X$, which is usually satisfied for Gaussian process; see \cite{shang2015nonparametric,yuan2010reproducing} for details. A6 is required for the asymptotic normality property of our test statistics. This assumption can be easily satisfied according to Proposition 4.2 in \cite{shang2015nonparametric}. A7 gives the desired order of the bandwidth $h$ for kernel estimation to guarantee the asymptotic properties.

Now we consider the general form of the underlying model:
\be
\label{general_model}
Y=\la X,\beta\ra+\delta_n\ell(X)+\eta,
\ee
where $E(\eta\mid X)=0$ and $\ell(\cdot)$ is a non-constant function.
When $\delta_n\equiv 0$, \eqref{general_model} refers to models under null hypothesis. It can also represent global and local alternatives when $\delta_n\equiv C\neq0$ and $\delta_n\rr0$, respectively. The following lemmas and theorems indicate the asymptotic properties of $\hat{\beta}_{n,M,\lambda}$, $\hat{q}$ and $T_n$ with different underlying models.

\subsection{Two-stage estimator}

Let $\hat{\beta}_{n,\lambda}$ be the regularized estimator based on fully observed functional data
\be
\hat{\beta}_{n,\lambda}=\underset{\beta \in \mathcal{H}}{\arg \min }\left\{\frac{1}{n} \sum_{i=1}^{n}\left[Y_{i}-\la X_i,\beta\ra \right]^{2}+\f{\lambda}{2} J(\beta,\beta)\right\}.
\ee
As has been proved in \cite{yuan2010reproducing}, if we take $\lambda=n^{-2k/(2k+1)}$, where $k$ is defined in Remark \ref{rem2}, then $\hat{\beta}_{n,\lambda}$ reaches optimal convergence rate $\|\hat{\beta}_{n,\lambda}-\beta^*\|_{L^2}=O_p(n^{-k/(2k+1)})$.  With this property, the consistency of the two-stage estimator under null and global alternatives can be obtained from the theorem below.

\begin{theorem}
\label{thm_estimator}
Let Assumptions A1 through A7 be satisfied, then under null and global alternatives, we have $\|\hat{\beta}_{n,M,\lambda}-\beta^*\|_{L^2}\leq O_p(n^{-k/(2k+1)})+O(M^{-r})$.
\end{theorem}

The functional principal component analysis (FPCA) based method can achieve a faster convergence rate as $n^{-1/2}$. However, the price it need to pay is the complicated inference results and intractable theoretical analysis. The convergence rate of the two-stage estimator consists of two parts. The first term is related to the estimation procedure, which is proved to be optimal under the RKHS framework. The second term reflects the smoothing procedure for discretely observed data. When the number of observations $M$ on a curve becomes large enough, its influence on the estimator will decay quickly. This provides theoretical results for the minimal points we should sample on a curve in practice.
We complete this subsection with the following theorem, which indicates the consistency of $\hat{\beta}_{n,M,\lambda}$ under local alternatives.

\begin{theorem}
\label{thm_estimator_local}
Let Assumptions A1 through A7 be satisfied, under the local alternatives we have
$$
\|\hat{\beta}_{n,M,\lambda}-\beta^*\|_{L^2}\leq O_p(n^{-k/(2k+1)})+O(M^{-r})+O_p(\delta_n)
$$
\end{theorem}

\subsection{Indicative dimension}

It's easy to derive the convergence rate of estimated indicative operator $\widehat{M}^{cc}$ under null and global alternatives.
\begin{theorem}
\label{thm_Mcc}
Let Assumptions A1 through A7 be satisfied, then $\widehat{M}^{c c}$ satisfies:
\begin{itemize}

\item[(1)] under the null hypothesis, $\|\widehat{M}^{cc}-M^{cc}\|=O_p(n^{-2k/(2k+1)})+O(M^{-2r})$;

\item[(2)] under the global alternatives, $\|\widehat{M}^{cc}-M^{cc}\|=O_p(n^{-k/(2k+1)})+O(M^{-r})$.
\end{itemize}
\end{theorem}

The convergence rate of estimated indicative operator $\widehat{M}^{cc}$ implies the properties of its eigenvalues. It paves the way to analyze the asymptotic behavior of $\hat{q}$ with TDRR method. The following theorem states the consistency of $\hat{q}$, which is used to indicate the underlying model.

\begin{theorem}
\label{thm_q1}
Let Assumptions A1 through A7 be satisfied, we have
\begin{itemize}
\item[(1)] if $c_{1 n} \rightarrow 0, c_{2 n} \rightarrow 0$ and $c_{1 n} c_{2 n} /(n^{-2k/(2k+1)}+M^{-2r})^2\rightarrow \infty,$ then under the null hypothesis, $\mathbb{P}(\hat{q}=0) \rightarrow 1$.

\item[(2)] if $c_{1 n} \rightarrow 0, c_{2 n} \rightarrow 0$ and $c_{1 n} c_{2 n} /(n^{-2k/(2k+1)}+M^{-2r}) \rightarrow \infty,$ then under the alternative hypothesis, $\mathbb{P}(\hat{q}>0) \rightarrow 1$.
\end{itemize}
\end{theorem}

From Theorem \ref{thm_q1}, we can see that $\hat{q}$ indeed has the ability to indicate the type of underlying model. Our hybrid test statistics $T_n$ will degenerate to $V_0$ and $V_1$ under null and global alternatives respectively, which means $T_n$ will share their strengths while avoiding their shortcomings. Also we should be aware that the assumptions of two ridges $c_{1n}$ and $c_{2n}$ listed in Theorem \ref{thm_q1} is for theoretical analysis. In practice, they will be selected by data-driven approach to be adaptive to different underlying models. We now give the asymptotic behavior of $\hat{q}$ under the local alternatives.

\begin{theorem}
\label{thm_q2}
Let Assumptions A1 through A7 be satisfied, then under the local alternatives, suppose that $\delta_n=n^{-\alpha}$, then we have
\begin{itemize}
\item[(1)] when $\alpha\geq{k}/(2k+1)$, let $c_{1 n} \rightarrow 0, c_{2 n} \rightarrow 0$ and $c_{1 n} c_{2 n}/(n^{-2k/(2k+1)}+M^{-2r})^2 \rightarrow \infty$, then $\mathbb{P}(\hat{q}=0) \rightarrow 1$.

\item[(2)] when $0<\alpha<{k}/(2k+1)$, let $c_{1 n}=o\left(\delta_{n}^{4}\right)$ and $c_{1 n} c_{2 n} /(n^{-8k/(2k+1)}+M^{-4r}) \rightarrow \infty$, then $\mathbb{P}(\hat{q}=q>0) \rightarrow 1$.
\end{itemize}
\end{theorem}

When the order of deviation term $\delta_n$ is close to $n^{-k/(2k+1)}$, the estimation $\hat{q}$ will jump from $0$ to some positive integer. This property implies that it can detect local alternatives with a deviation term slower than $n^{-k/(2k+1)}$. Unlike the $n^{-1/2}$ threshold shown in \cite{li2021adaptive}, there is a shrinkage of critical order of $\delta_n$ for functional data, which is the price we have to pay to use reproducing kernel based estimator. However, since $k=m+s+1$, we can set a larger $m$ to get the critical order closer to $n^{-1/2}$. Finally, we should be aware that these results only have the theoretical meaning, unless we have prior information on the closeness of local alternatives to the null.

\subsection{Test statistics}

With the results shown above, we now discuss the asymptotic properties of the proposed test in detail under the null, global alternative and local alternative hypothesis.

\begin{theorem}
\label{thm6}
Let Assumptions A1 through A7 be satisfied, if $Mn^{-1/2r}\rr\infty$, then
\begin{itemize}
\item[(1)] under the null hypothesis, $T_n\dr \chi_1^2$.

\item[(2)] under the global alternative hypothesis, $T_n/\gamma_{n,M}$ converges to a constant $\mu>0$.
\end{itemize}
\end{theorem}

Theorem \ref{thm6} indicates that the hybrid test statistics $T_n$ will converge in distribution to $\chi_1^2$ under the null hypothesis. Therefore, the critical value of the model checking test can be easily determined without using any resampling techniques, releasing the computation burden. Under the global alternatives, the proposed test will diverge to infinity at order $\gamma_{n,M}=O_p(n^{2k/(2k+1)})$. Suppose all the assumptions mention above hold, we can conclude that the convergence rate of $V_0^2$ and $V_1$ under the null hypothesis are $n^{-2k/(2k+1)}$ and $n^{-1}h^{1/2}$, respectively. According to the non-parametric kernel theory, the optimal choice of the bandwidth $h$ in $V_1$ is $n^{-2/5}$, then the order of $V_1$ is $n^{-4/5}$ under the null hypothesis. In practice, we usually take the order of Sobolev space $m\geq2$, and thus $k\geq3$, which means $V_0$ converges faster than $V_1$. Meanwhile, $\gamma_{n,M}V_1$ will diverge faster than $nh^{-1/2}V_1$ under the global alternatives,  Therefore, the hybrid test is more powerful than simply use any one of $V_0$ and $V_1$. We also obtain the asymptotic distribution of $T_n$ under the local alternatives.

\begin{theorem}
\label{thm8}
Let Assumptions A1 through A7 be satisfied, then under the local alternatives, let $\delta_n=n^{-\alpha}$, if $Mn^{-1/2r}\rr\infty$, we have
\begin{itemize}
\item[(1)] if $\alpha>k/(2k+1)$, $T_n\dr\chi_{1}^2$.

\item[(2)] if $\alpha=k/(2k+1)$, $T_n\dr \chi_{1}^2(\mu_0)$, where $\chi_{1}^2(\mu_0)$ is a chi-squared distribution with one degree of freedom and noncentrality parameter $\mu_0\neq0$.

\item[(3)] if $0<\alpha<k/(2k+1)$, and

\begin{itemize}
\item[(a)]
$n^{1 / 2} h^{1/ 4} \delta_{n} \rightarrow 0$, then $T_{n} / h^{-1 / 2}$ converges in distribution to $N(0,\Sigma)$, where
$$
\Sigma=2 \int K^{2}(u) \mathrm{d} u \int\left\{\sigma^{2}(\mathrm{z})\right\}^{2} f^{2}(\mathrm{z}) \mathrm{d} \mathrm{z},
$$
where $Z=\la \beta_0,X\ra, \sigma^{2}(z)=E\left(\epsilon^{2} \mid Z=z\right)$.

\item[(b)] $\delta_{n}=n^{-1 / 2} h^{-1 / 4}$, then $T_{n} / h^{-1 / 2}$ converges in distribution to $N(E(\ell^{2} f),\Sigma)$, where $f(z)$ be the probability density function of $Z=\la \beta_0, X\ra$.

\item[(c)] $n^{1 / 2} h^{1 / 4} \delta_{n} \rightarrow \infty$, then $T_{n} /\left(n \delta_{n}^{2}\right)$ converges in probability to $E\left(\ell^{2} f\right)$.
\end{itemize}
\end{itemize}
\end{theorem}

Theorem \ref{thm8} states that the hybrid test can detect the alternatives with a deviation term slower than or equal to $n^{-k/(2k+1)}$. When $\delta_n=n^{-k/(2k+1)}$, the estimated indicative dimension $\hat{q}$ will be zero with probability going to $1$ according to Theorem \ref{thm_q2}, reducing $T_n$ to $\gamma_{n,M}V_0^2$. The influence of the deviation term will then be reflected in the asymptotic property of $\gamma_{n,M}^{1/2}V_0$, making it a non-central normal distribution. The third part gives a full picture to show what rate of divergence we can achieve when $\delta_n$ is slower than $n^{-k/(2k+1)}$. Until now, we have systematically studied the asymptotic behaviors of $T_n$ with different underlying models. The results demonstrate the expected merits of the hybrid test inherited from moment-based and conditional moment-based tests. Finally, we note that the condition $Mn^{-1/2r}\rr\infty$ shown in above three theorems implies that if $M=Cn^{1/2r}$ for some large enough $C$, the impact of discrete observations will be eliminated. In practice, we would suggest that $C>20$.


\section{Numerical Studies}

\label{numerical}

In this section, we perform simulation studies using nine scenarios defined in \cite{cuesta2019goodness} and compare their powers. The different data generating processes are encoded as follows. For the $k$-th simulation scenario $S_k$, with slope function $\beta_k$, the deviation from $H_0$ is measured by a deviation coefficient $\delta_d$, with $\delta_{0}=0$ and $\delta_{d}>0$ for $d=1,2$. Then, we denote $H_{k, d}$ the data generation from
$$
Y=\left\langle X, \beta_{k}\right\rangle+\delta_{d} \ell_{j}(X)+\eta,
$$
where $j\in\{1,2,3\}$ and the deviations from the linear model are defined by the nonlinear terms $\ell_{1}(X):=\|X\|, \ell_{2}(X):=25 \int_{0}^{1} \int_{0}^{1} \sin (2 \pi t s) s(1-s) t(1-t) X(s) X(t) \mathrm{d} s \mathrm{~d} t$, and $\ell_{3}(X):=\left\langle e^{-X}, X^{2}\right\rangle .$ The error term $\eta$ follows a centered normal distribution $\mathcal{N}\left(0, \sigma^{2}\right)$, where $\sigma^{2}$ is chosen such that, under $H_{0}, R^{2}={\operatorname{Var}[\langle\mathrm{X}, \beta\rangle]}/({\operatorname{Var}[\la\mathrm{X}, \beta\rangle]+\sigma^{2}})=0.95$. The description of the simulation scenarios is given in Table \ref{tab1}. We select five types of functional processes $X(t)$, all of them defined on $[0,1]$:

\textbf{BM.} Brownian motion, denoted by $\mathbf{B}$, with eigenfunctions $\psi_{j}(t):=\sqrt{2} \sin ((j-0.5) \pi t)$, $j \geq 1$, will be generated by $X(t)=\sum_{j=1}^{100} Z_j\psi_{j}(t)/((j-0.5)\pi)$, where $Z_j$ are i.i.d. standard normal random variables.

\textbf{HHN.} The functional process considered in \cite{hall2006properties}, given by $X(t)=$ $\sum_{j=1}^{20} \xi_{j} \phi_{j}(t)$, where $\phi_{j}(t):=\sqrt{2} \cos (j \pi t)$ and $\xi_{j}$ are i.i.d. random variables distributed as $\mathcal{N}(0, j^{-2 l})$, with $l=1,2$.

\textbf{BB.} Brownian bridge, defined as $X(t)=\mathbf{B}(t)-t \mathbf{B}(1)$.

\textbf{OU}. Ornstein-Uhlenbeck process, defined as the zero-mean Gaussian process with covariance given by $\operatorname{Cov}[X(s), X(t)]=\sigma^{2}/(2 \alpha) e^{-\alpha(s+t)}(e^{2 \alpha \min (s, t)}-1)$. We consider $\alpha=1/3, \sigma=1$, and $X(0) \sim$ $\mathcal{N}(0, \sigma^{2}/(2 \alpha))$. It can be generate by $X(t)=(\sigma/\sqrt{2\alpha})e^{-\alpha t}\mathbf{B}(e^{2\alpha t})$.

\textbf{GBM.} Geometric Brownian motion, defined as $X(t)=s_{0} \exp \{(\mu-\sigma^{2}/{2}) t+\sigma \mathbf{B}(t)\}$. We consider $\sigma=1, \mu=0.5$, and $s_{0}=2$

\begin{table*}
    \centering
    \caption{Simulation scenarios and deviations from the null hypothesis}
    \label{tab1}
    \scalebox{1}{\begin{tabular}{c|l|l|l}
        \hline \hline Scenario & Coefficient $\beta(t)$ & Process $X$ & Deviation \\
        \hline $\mathrm{S} 1$ & $(2 \psi_{1}(t)+4 \psi_{2}(t)+5 \psi_{3}(t)) / \sqrt{2}$ & $\mathrm{BM}$ & $\ell_{1}, \delta=(0, 0.25, 0.75)^{\prime}$ \\[0.5em]
        \hline $\mathrm{S} 2$ & $(2 \tilde{\psi}_{1}(t)+4 \tilde{\psi}_{2}(t)+5 \tilde{\psi}_{3}(t)) / \sqrt{2}$ & $\mathrm{BB}$ & $\ell_{2}, \delta=(0,-2,-7.5)^{\prime}$ \\[0.5em]
        \hline $\mathrm{S} 3$ & $(2 \psi_{2}(t)+4 \psi_{3}(t)+5 \psi_{7}(t)) / \sqrt{2}$ & $\mathrm{BM}$ & $\ell_{1}, \delta=(0,-0.2,-0.5)^{\prime}$ \\[0.5em]
        \hline $\mathrm{S} 4$ & $\sum_{j=1}^{20} 2^{3 / 2}(-1)^{j} j^{-2} \phi_{j}(t)$ & $\mathrm{HHN}(l=1)$ & $\ell_{2}, \delta=(0,-1,-3)^{\prime}$ \\[0.5em]
        \hline $\mathrm{S} 5$ & $\sum_{j=1}^{20} 2^{3 / 2}(-1)^{j} j^{-2} \phi_{j}(t)$ & $\mathrm{HHN}(l=2)$ & $\ell_{2}, \delta=(0,-1,-3)^{\prime}$ \\[0.5em]
        \hline $\mathrm{S} 6$ & $\log (15 t^{2}+10)+\cos (4 \pi t)$ & $\mathrm{BM}$ & $\ell_{1}, \delta=(0, 0.2, 1)^{\prime}$ \\[0.5em]
        \hline $\mathrm{S} 7$ & $\sin (2 \pi t)-\cos (2 \pi t)$ & $\mathrm{OU}$ & $\ell_{2}, \delta=(0,-0.25,-1)^{\prime}$ \\[0.5em]
        \hline $\mathrm{S} 8$ & $t-(t-0.75)^{2}$ & $\mathrm{OU}$ & $ \ell_{3}, \delta=(0,-0.01,-0.1)^{\prime}$ \\[0.5em]
        \hline $\mathrm{S} 9$ & $\pi^{2}(t^{2}-1/3)$ & $\mathrm{GBM}$ & $ \ell_{3}, \delta=(0, 0.5, 2.5)^{\prime}$ \\
        \hline \hline
    \end{tabular}}
\end{table*}

\subsection{Data-driven ridge selection}
As an important component of hybrid test, indicative dimension needs to be well estimated so that $\hat{q}$ has the desired properties. The ridges $c_{1n}$ and $c_{2n}$ in TDRR method should be carefully selected to satisfy conditions presented in Theorem \ref{thm_q1} and Theorem \ref{thm_q2}. Existing ridge selection methods \cite{guo2016model,li2021adaptive} usually use pre-determined fix numbers based on numerical experiences, which is not adaptive to different underlying models. For instance, the mean and variance of the nine simulation scenarios we considered vary a lot. The fixed ridges cannot achieve satisfying performance in all scenarios, which urges us to develop a data-driven ridge selection method. In the subsection, we will illustrate the construction of $c_{1n}$ and $c_{2n}$ in a data-driven way.

The art of ridge selection should satisfy two requirements simultaneously, (r1): it should achieve the asymptotic properties stated in Theorem \ref{thm_q1} and Theorem \ref{thm_q2}; (r2): it should be adaptive to different process $X(t)$ and slope function $\beta_0(t)$. To satisfy (r1), one good choice is to set $c_{1n}=c_{2n}=2\hat{s}_1$, where $\hat{s}_1=\hat{\lambda}_1/(\hat{\lambda}_1+1)$ and $\hat{\lambda}_1$ is the largest eigenvalue of $\wh{M}_{cc}$ under null hypothesis. This will guarantee $\hat{q}=0$ under the null hypothesis using TDRR method and $\hat{q}>0$ under alternatives when the deviation is large enough. However, we have no prior information on our underlying model, so we can't tell whether $\wh{M}^{cc}$ is estimated from the model under the null hypothesis.

To address this problem, we note that $\wh{M}^{cc}$ only depends on the smoothed functional data $\wh{X}_i(t)$ and remainders $\hat{\epsilon}_i$. If we assume our data comes from a FLM with error term $\eta$ such that $E(\eta\mid X)=0$ and $\op{Var}(\eta)=\sigma^2$, let
\be
\label{M_null}
\widehat{M}_{null}=\hat{E}(\eta X) \otimes \hat{E}\left(\eta X\right)+\hat{H} \hat{H}
\ee
where $\hat{E}(\eta X)= (\sum_{j=1}^{n} \eta_{j} \widehat{X}_{j})/n$, $\hat{H}= (\sum_{j=1}^{n} \eta_{j} \widehat{X}_{j}\otimes \widehat{X}_{j})/n$ and $\eta_j$ are drawn i.i.d. from $N(0,\sigma^2)$. Then the eigenvalues of $\widehat{M}_{null}$ should be close to the eigenvalues of $\wh{M}_{cc}$ under the null hypothesis. Replicate calculating \eqref{M_null} for $B$ times we get $\widehat{M}_{null}^{(1)},\ldots,\widehat{M}_{null}^{(B)}$, then we average them to mitigate the randomness of sampling $\eta_j$ by setting $\wh{M}^*=(\sum_{i=1}^B\widehat{M}_{null}^{(i)})/B$. In practice, we take $B=100$ and $\sigma^2$ is estimated by the variance of residuals $\hat{\epsilon}_i$. The ridge will be chosen as $c_{1n}=c_{2n}=2\hat{s}_1^*$, where $\hat{s}_1^*=\hat{\lambda}_1^*/(\hat{\lambda}^*_1+1)$ and $\hat{\lambda}^*_1$ is the largest eigenvalue of $\wh{M}^*$. The trick of our approach is to replace the remainders under the null hypothesis by i.i.d. samples from normal distributed variables with the same variance. Although we don't know the exact distribution information of the error term under null hypothesis, numerical experiments shows that the difference of $\hat{s}_1^*$ and $\hat{s}_1$ under the null hypothesis can be ignored since $\eta_j$ shares the same mean and variance with $\eta$. Theoretical analysis and discussion of this data-driven selection procedure need further study.

In practice, we also suggest setting a support order to deal with small variance models. When $\sigma^2$, the variance of the error term, is small, the largest eigenvalues $\hat{\lambda}^*_1$ would also be small, which makes the TDRR method extremely sensitive to the deviations. In this case, $\hat{q}$ can easily jump to a positive value even under the null hypothesis. To ensure the robustness of our criteria, it is recommended to set $c_{1n}=c_{2n}=\max\{2\hat{s}_1^*,n^{-0.5}\}$.

\subsection{Simulation results}

In this subsection, the numerical experiments are conducted on nine scenarios defined above using our hybrid test statistics $T_n$. The results will be compared to the methods proposed in \cite{garcia2014goodness} and \cite{cuesta2019goodness}. The sample size is chosen to be $n=100,250$. Throughout the simulations, the stochastic process $X(t)$ is observed at $M=30$ equidistant points. The weight function $w(X)$ is chosen as $w(X)=0.01\|X\|$ and the ridge $c_{1n}$ and $c_{2n}$ are chosen from the data driven way mentioned above. The significance level is $\alpha=0.05$. The bandwidth $h$ is chosen as optimal value $n^{-0.4}$ according to non-parametric kernel theory. The penalty parameter $\lambda$ is selected by the generalized cross validation (GCV) illustrated in \cite{yuan2010reproducing}. We replicate the experiments for 1000 times in each setting to calculate the empirical size or empirical power for our test statistics. The performances of $CvM_3$, $KS_3$ and $PCvM$ methods are borrowed from \cite{cuesta2019goodness}. The results are listed in Table \ref{tab2}.

We find that the power of $T_n$ is competitive in all scenarios, especially in $S_6$ and $S_8$ when $n=250$.  In these two scenarios, the slope function $\beta_0$ is not a linear combination of eigenfunctions of the covariance function $C(s,t)$. Therefore, the functional principal component analysis based estimation methods used in \cite{cuesta2019goodness} will perform badly. Throughout the nine scenarios, our hybrid test $T_n$ has higher powers than $CvM_3$ and $KS_3$ when the deviation term is relatively small. Specifically, $T_n$ enjoys highest rejection rate in $H_{1,1}$, $H_{2,1}$, $H_{4,1}$, $H_{5,1}$ $H_{7,1}$ and $H_{9,1}$ among other methods. The results indicate that $T_n$ has advantages in detecting local alternatives. For larger deviations, the rejection rate of $T_n$ also shows its superiority to $CvM_3$ and $KS_3$, and is comparable to $PCvM$. All results above are evidences of the fact that $T_n$ shares the merits from both $V_0$ and $V_1$ and thus become powerful under the local and global alternatives while having a chi-square distribution under the null hypothesis.

\begin{table}[h]
    \centering
    \caption{Empirical sizes and powers in percentages for nine scenarios}
    \label{tab2}
    \begin{tabular}{c|cccc|cccc}
        \toprule
        & \multicolumn{4}{c}{n=100} & \multicolumn{4}{c}{n=250}\\ \cmidrule(lr){2-5} \cmidrule(lr){6-9}
        $H_{k,\delta}$ & $CvM_3$ & $KS_3$ & $PCvM$ & $T_n$ & $CvM_3$ & $KS_3$ & $PCvM$ & $T_n$\\
        \midrule
        $H_{1,0}$ & 3.9 & 4.6 & 4.8 & 5.4 & 3.9 & 4.2 & 4.9 & 5.1\\
        $H_{2,0}$ & 4.6 & 5.1 & 3.6 & 5.7 & 4.8 & 5.4 & 4.7 & 4.9\\
        $H_{3,0}$ & 4.9 & 6.0 & 5.7 & 3.8 & 4.1 & 4.7 & 5.3 & 4.4\\
        $H_{4,0}$ & 4.4 & 5.0 & 4.6 & 5.4 & 5.2 & 5.9 & 4.9 & 4.7\\
        $H_{5,0}$ & 4.0 & 4.3 & 4.9 & 5.6 & 4.2 & 4.0 & 5.0 & 4.7\\
        $H_{6,0}$ & 4.3 & 4.9 & 5.2 & 5.3 & 4.3 & 5.0 & 4.8 & 5.2\\
        $H_{7,0}$ & 3.9 & 4.7 & 5.1 & 6.1 & 4.1 & 4.7 & 5.2 & 4.5\\
        $H_{8,0}$ & 3.5 & 3.7 & 4.9 & 4.6 & 3.9 & 4.3 & 5.1 & 5.1\\
        $H_{9,0}$ & 4.8 & 4.7 & 6.1 & 6.0 & 4.4 & 4.8 & 5.9 & 5.4\\
        \midrule
        $H_{1,1}$ & 59.4 & 45.0 & 69.9 & 98.5 & 96.3 & 90.3 & 98.4 & 100\\
        $H_{2,1}$ & 98.5 & 95.7 & 99.2 & 99.5 & 100 & 100 & 100 & 100\\
        $H_{3,1}$ & 97.6 & 93.0 & 99.2 & 98.8 & 100 & 100 & 100 & 100\\
        $H_{4,1}$ & 35.7 & 26.8 & 43.6 & 48.9 & 81.8 & 67.7 & 88.6 & 82.9\\
        $H_{5,1}$ & 43.1 & 31.8 & 49.9 & 51.3 & 87.9 & 75.3 & 91.5 & 89.1\\
        $H_{6,1}$ & 22.2 & 17.0 & 27.9 & 23.7 & 57.0 & 43.0 & 66.9 & 75.6\\
        $H_{7,1}$ & 99.9 & 99.8 & 99.9 & 99.9 & 100 & 100 & 100 & 100\\
        $H_{8,1}$ & 74.8 & 50.3 & 74.7 & 74.4 & 88.3 & 76.0 & 87.7 & 97.3\\
        $H_{9,1}$ & 9.2 & 8.9 & 12.1 & 13.4 & 17.9 & 16.4 & 22.3 & 20.9\\
        \midrule
        $H_{1,2}$ & 100 & 99.9 & 100 & 100 & 100 & 100 & 100 & 100\\
        $H_{2,2}$ & 99.8 & 99.5 & 99.9 & 100 & 100 & 100 & 100 & 100\\
        $H_{3,2}$ & 100 & 100 & 100 & 100 & 100 & 100 & 100 & 100\\
        $H_{4,2}$ & 96.4 & 92.0 & 98.2 & 97.8 & 99.9 & 99.9 & 100 & 99.7\\
        $H_{5,2}$ & 98.9 & 97.0 & 99.1 & 98.1 & 100 & 100 & 100 & 99.6\\
        $H_{6,2}$ & 100 & 99.8 & 100 & 99.9 & 100 & 100 & 100 & 100\\
        $H_{7,2}$ & 99.9 & 99.9 & 99.9 & 100 & 100 & 100 & 100 & 100\\
        $H_{8,2}$ & 76.5 & 45.8 & 78.2 & 99.6 & 88.0 & 73.0 & 88.9 & 100\\
        $H_{9,2}$ & 90.5 & 85.9 & 93.9 & 83.1 & 100 & 100 & 100 & 96.1\\

        \bottomrule
    \end{tabular}

\end{table}

\subsubsection{Hybrid effect}
In this part, we examine the expected properties of the hybrid tests through comparisons of their components. The asymptotic distributions of $V_0$ and $V_1$ under the null hypothesis and be easily obtained from the proofs of Theorem \ref{thm6} and Theorem \ref{thm8}. We will demonstrate that the hybrid test $T_n$ is more powerful than simply use any one of $V_0$ and $V_1$. Let $Q_0$ be the percentage of the indicative dimension that is estimated to be zero, that is $P(\hat{q}=0)$, which will be used to verify the conclusions in Theorem \ref{thm_q1} and illustrate the unique property of $T_n$.

Consider the model as $Y=\left\langle X, \beta_0\right\rangle+\delta\ell(X)+\eta$, where $X$ is the Brownian motion. The slope function $\beta_0=\sum_{j=1}^{20} 4(-1)^{j} j^{-2}\cos (j \pi t)$ and $\ell(X)= 0.25\sin{(\la X,X\ra)}$ and $\eta\sim \mathcal{N}(0,0.15^2)$. We set sample size $n=100$, number of observations $M=200$, order of Sobolev space $m=2$ and the shift coefficient $\delta$ will change from $0$ to $1$. The empirical sizes and powers are shown in Tables \ref{tab3}.

\begin{table}[h]
    \centering
    \caption{Empirical sizes and powers for components of the hybrid test.}
    \label{tab3}
    \setlength\tabcolsep{20pt}
    \begin{tabular}{ccccccc}
        \toprule
        $\delta$ & $CvM_3$ & $PCvM$ & $T_n$& $V_0$ & $V_1$ & $Q_0$ \\
        \midrule
        0   & 4.7  & 3.8  &  5.2 & 5.1  & 2.3  &  99.6 \\
        0.2   & 5.6  & 5.8  & 9.2  & 7.4  & 3.4  & 95.4  \\
        0.4   & 23.1  &  23.8 & 67.4  &  38.6 &  10.2 & 36.4  \\
        0.6   & 48.2  & 46.7  & 88.6  & 60.4  & 25.2  & 12.6  \\
        0.8   & 68.2  & 70.5  & 97.6  & 79.3  &  54.7 & 3.5  \\
        1   & 77.9  &  85.3 & 99.4  & 87.5  & 76.1  &  0.7 \\
        \bottomrule
    \end{tabular}
\end{table}

It manifests that the hybrid test is more powerful than either $V_0$ or $V_1$ alone. As the shift term becomes larger, the percentage $Q_0$ converges to zero as expected, which implies our data driven estimation procedure performs well. Assume the bandwidth $h$ satisfies {A7}, then the convergence rate of $V_0$ is faster than $V_1$. Since the standardizing factor $\gamma_{n,M}$ of $T_n$ is determined by $V_0$,  when the indicative dimension is estimated to be positive, $T_n$ will be larger than $nh^{1/2}V_1$ and therefore more powerful.

\subsubsection{Number of observations}
In the subsection, we investigate how the number of observations $M$ on a curve influences the power of $T_{n}$. As has been discussed in the asymptotic property part, the error from smoothing is vanishing with a scale of $M^{-r}$ when $M$ is large enough. That is, when $M$ exceeds some threshold, its increase will not help to improve the performance. On the other hand, if we do not sample enough observations on a curve, the power of $T_n$ will not increase as simple size $n$ increases. Our theorems provide guidance to balance these two folds and give an economical and efficient sampling criterion.

\begin{figure}[h]
\centering
\subfloat[]{\resizebox{200pt}{!}{\includegraphics[width=0.4\textwidth]{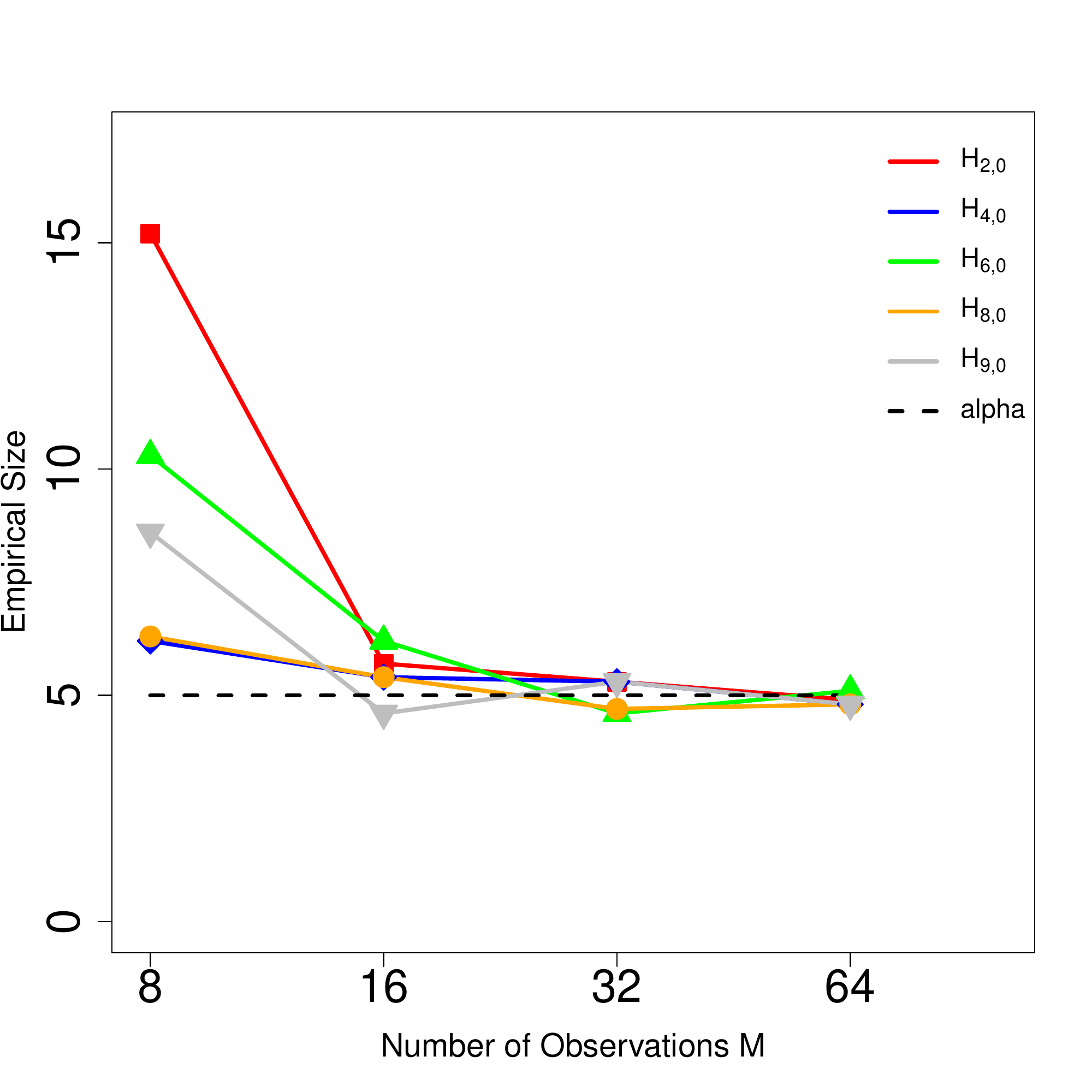}}}
\subfloat[]{\resizebox{200pt}{!}{\includegraphics[width=0.4\textwidth]{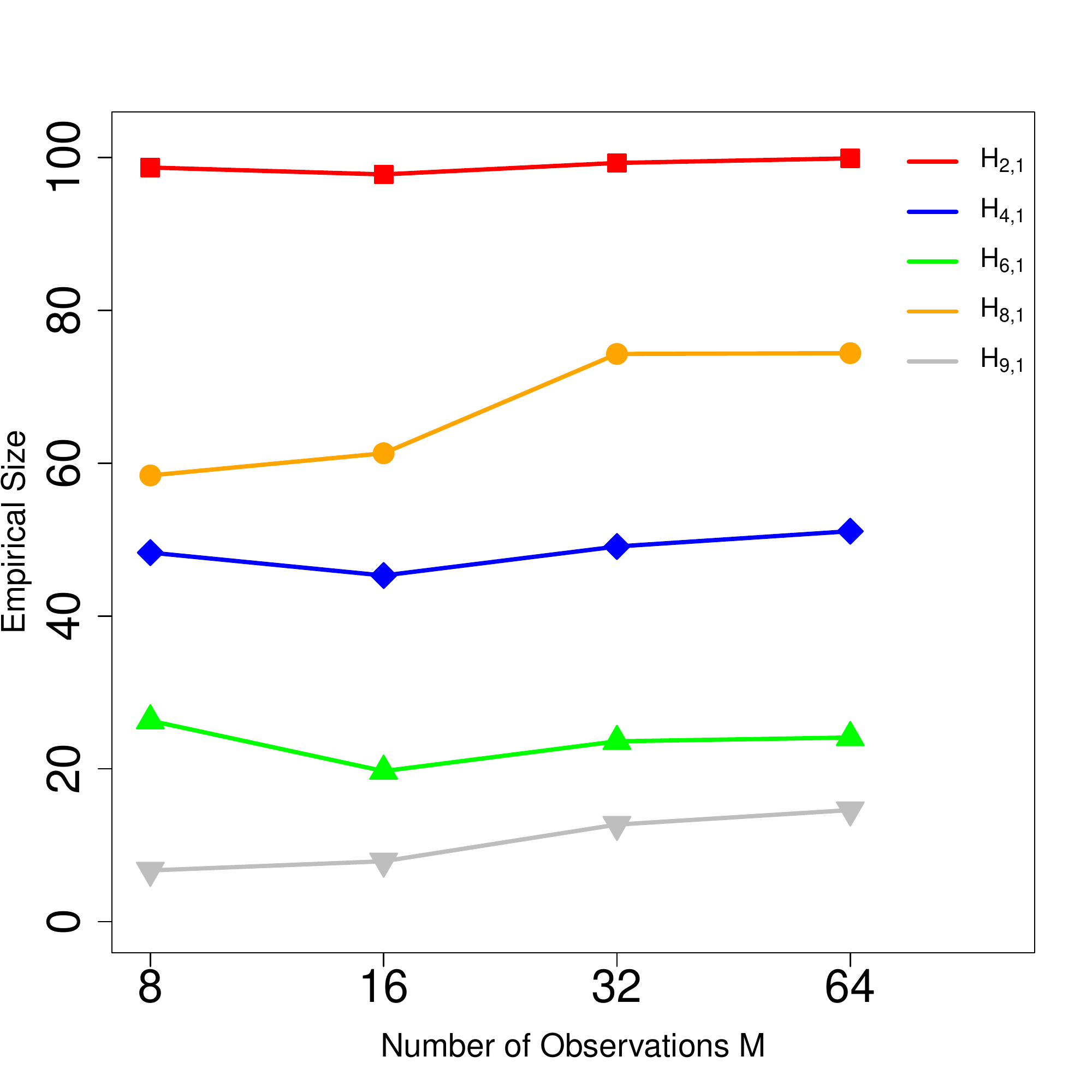}}}

\caption{Empirical sizes (a) and powers (b) versus number of observed points.}
\label{fig1}
\end{figure}

To verity these two patterns, we design the experiments as follows. In the first experiment, we set $M=2^k,\,k=3,4,5,6$, $n=100$, $\alpha=0.05$ and select $S_2$, $S_4$, $S_6$, $S_8$ and $S_9$ scenarios (five different stochastic processes $X(t)$). Other parameters are the same as previous subsection. The results are shown in Figure \ref{fig1}.

From Figure \ref{fig1}, we notice that $M=8$ is obviously not enough to describe a curve, because we can't control the empirical size to be close to $\alpha$. This makes the powers under alternatives not reliable. As $M$ increases, the empirical size converges to $\alpha$ and the empirical powers are improved gradually. When $M$ exceeds $32$, its effect is not significant any more, indicating that $M=32$ can capture enough information on a curve with support $[0,1]$.

\begin{figure}[h]
\centering
\resizebox{250pt}{!}{\includegraphics[width=0.4\textwidth]{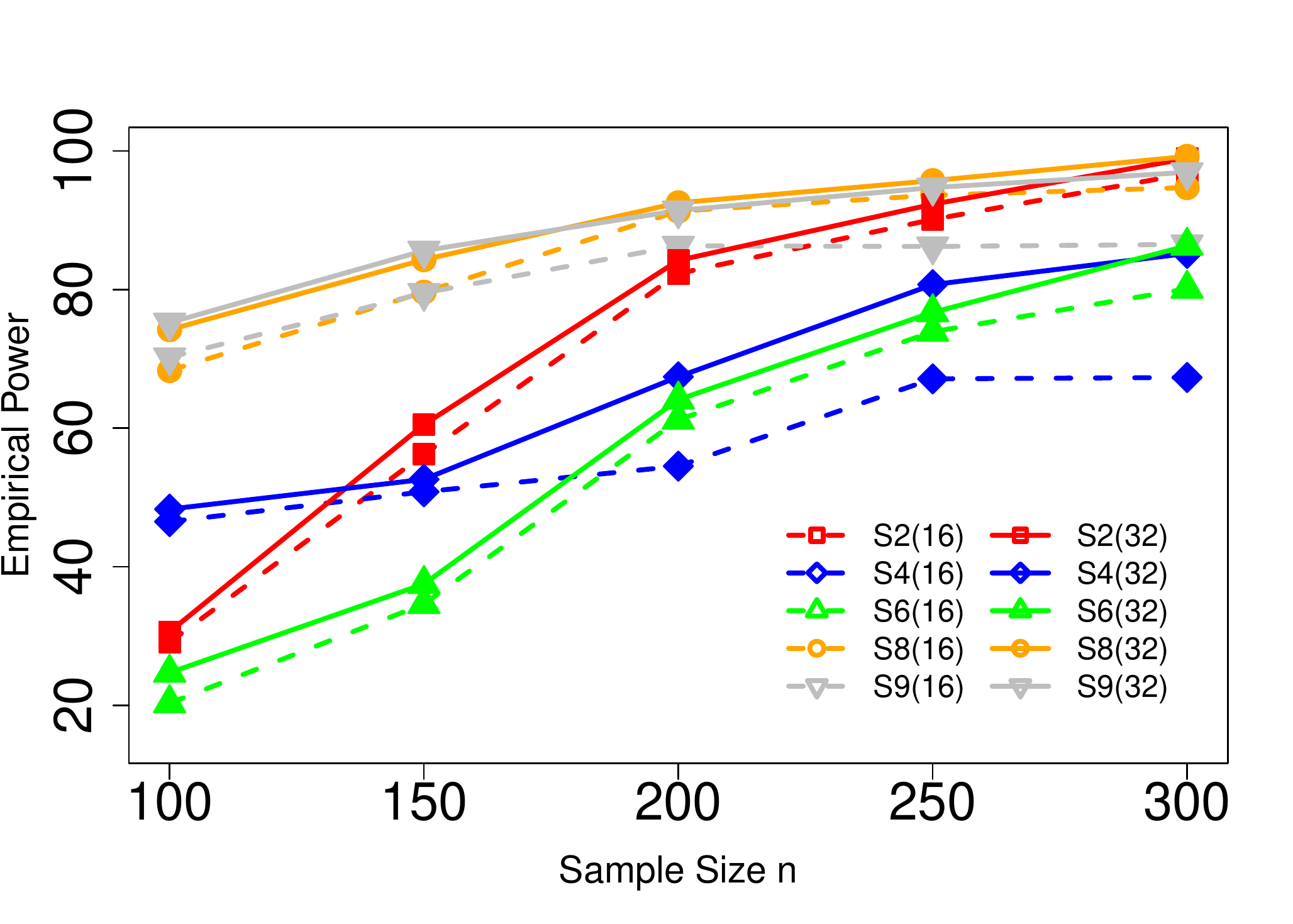}}

\caption{Empirical powers versus sample size for different number of observations $M$ in different scenario. Solid line for $M=32$ and dash line for $M=16$.}
\label{fig2}
\end{figure}

In the second experiment, we compare the trends of empirical powers under two settings $M=16$ and $M=32$ as sample size $n$ increase from $100$ to $300$. We still set $\alpha=0.05$ and select $S_2$, $S_4$, $S_6$, $S_8$ and $S_9$ scenarios. The deviation coefficient $\delta$ is set to be $-0.5$, $-1$, $0.2$, $-0.01$ and $2.5$ respectively. Other parameters are the same as above. The results are shown in Figure \ref{fig2}.

It can be concluded from Figure \ref{fig2} that for $S_2$, $S_6$ and $S_8$, solid lines and dash lines follow nearly the same trends with solid lines a little bit above the dash lines. In these three scenarios, the number of observations $M=16$ may already be enough to smooth the discrete functional data. If we choose $M=32$, it is more powerful in detecting alternatives, which coincides with the conclusion in the first experiment. In contrast, for $S_4$ and $S_9$, the power will gradually reach a bound and stop increasing even if we set a larger sample size. In this case, the term $M^{-1/2r}$ will dominate the convergence rate of two-stage estimator $\hat{\beta}_{n,M,\lambda}$ no matter how large the sample size $n$ is. It will influence the order of $V_0$ and $V_1$ and impair the power of the hybrid test. These two experiments well validate our theoretical analysis for the number of points observed on a curve, which will provide guidance for us to choose an optimal $M$ that balances the sample costs as well as the quality of functional data.

\subsection{Real Data Application}
\begin{figure*} [t!]
	\centering
	\subfloat[\label{fig:a}]{
		\resizebox{150pt}{!}{\includegraphics[width=0.4\textwidth]{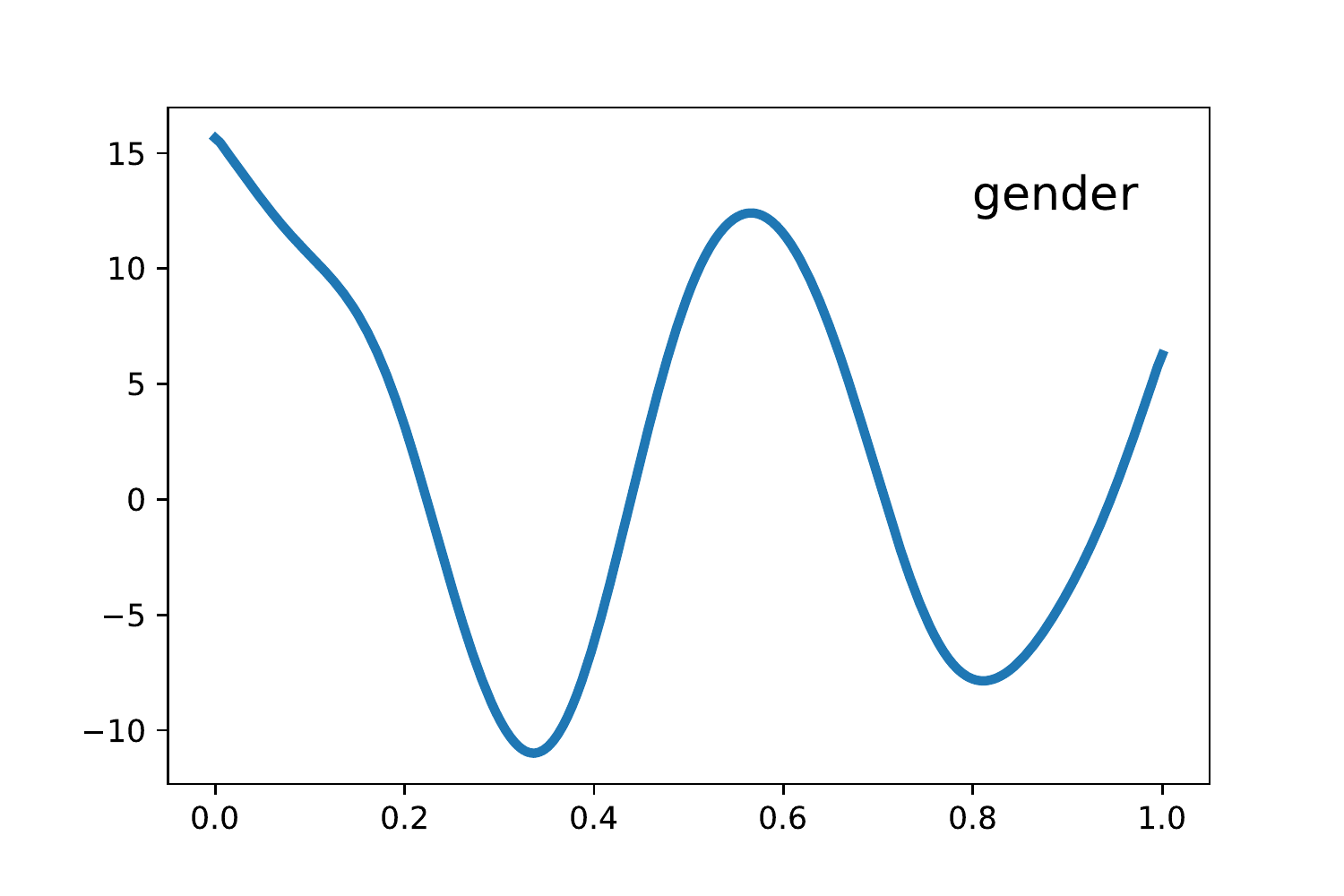}}}
	\subfloat[\label{fig:b}]{
		\resizebox{150pt}{!}{\includegraphics[width=0.4\textwidth]{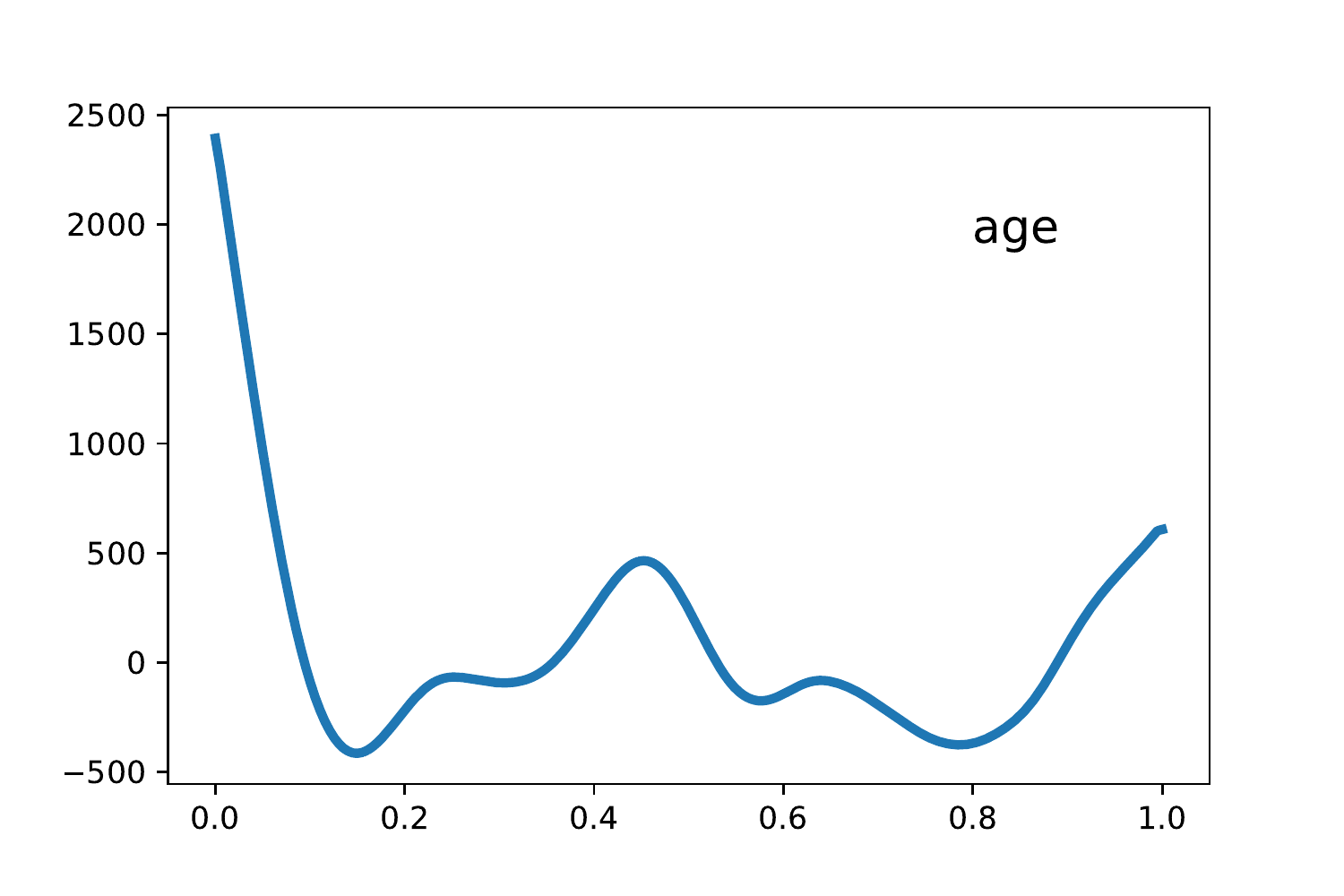}}}
	\\
	\subfloat[\label{fig:c}]{
		\resizebox{150pt}{!}{\includegraphics[width=0.4\textwidth]{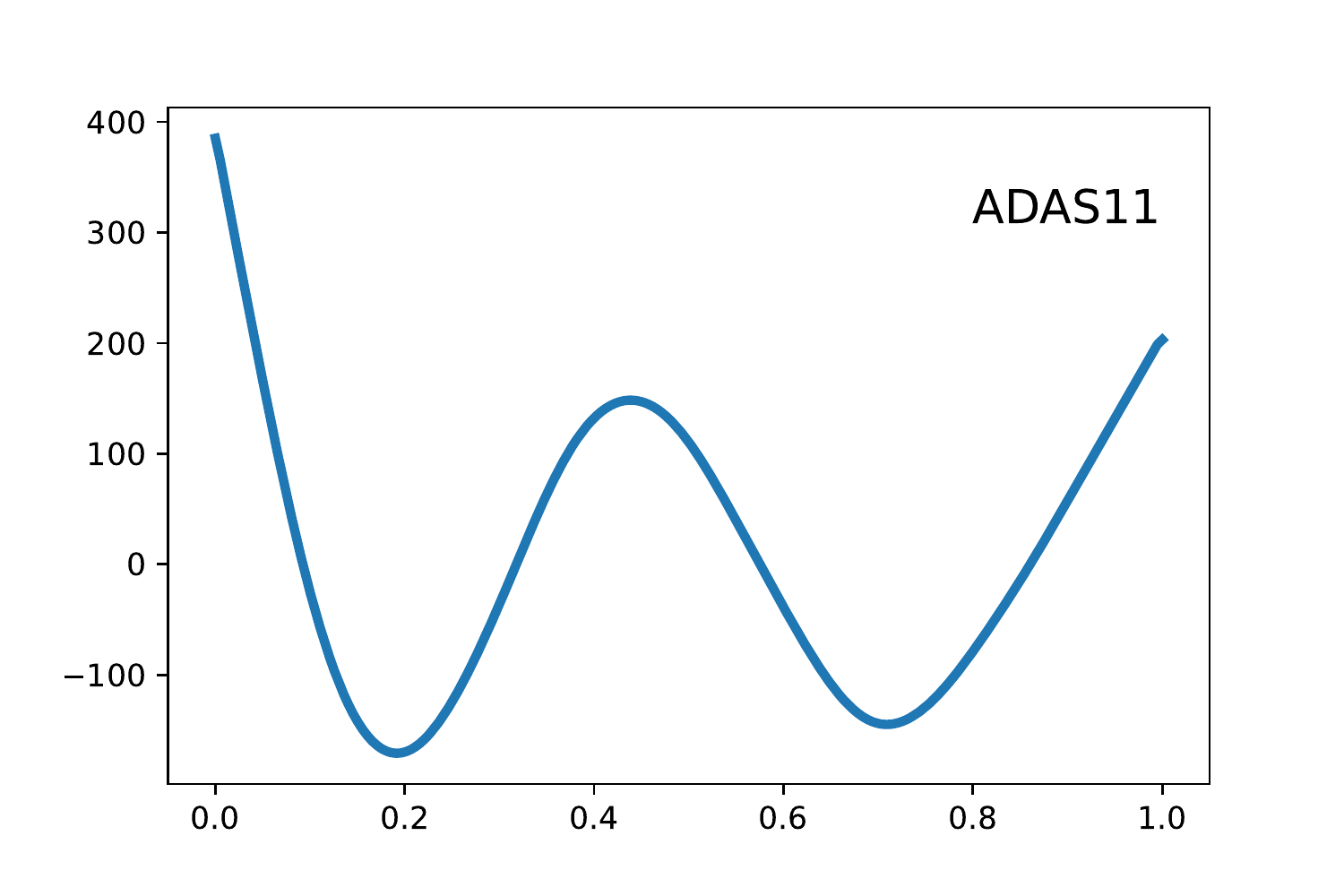}}}
	\subfloat[\label{fig:d}]{
	    \resizebox{150pt}{!}{\includegraphics[width=0.4\textwidth]{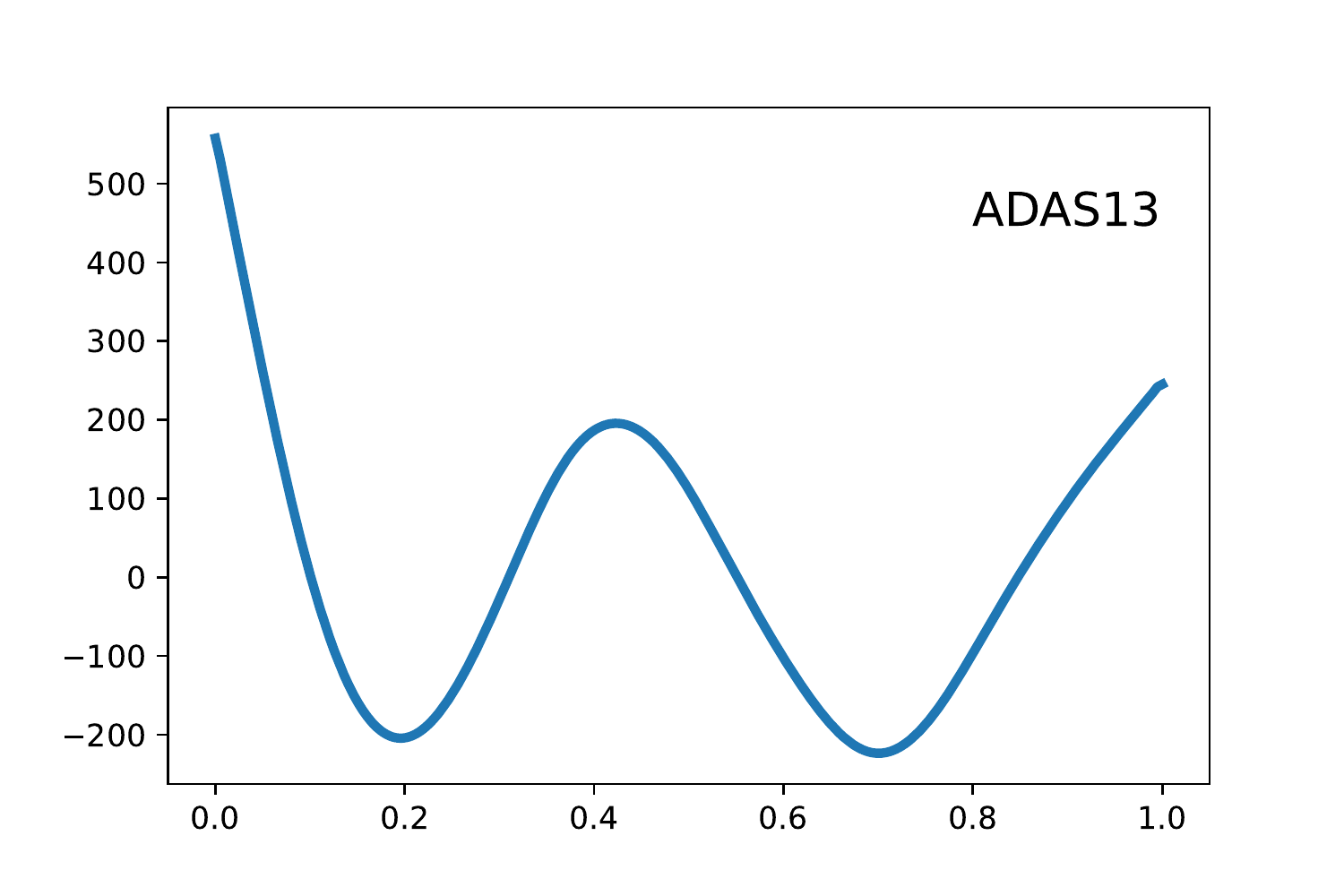}}}
	\caption{From (a) to (d): Estimated slope functions of gender, age, ADAS11 and ADAS13 for the ADNI DTI data.}
	\label{fig3}
\end{figure*}

We apply the hybrid test to three real datasets to examine whether FLM is sufficient to describe the data. First, we analyze the diffusion tensor imaging (DTI) data in the Alzheimer's Disease Neuroimaging Initiative (ADNI) study, which has also been studied in \cite{zhang2021high}.
In this study, there are 217 subjects in total, with four outliers. The outliers can be identified by the plot and will be eliminated before the test. The fractional anisotropy (FA), a scalar measure of the degree of anisotropy, along the corpus callosum (CC) with 83 equally spaced grid points can be regarded as the discretely observed functional variable $X(t)$ in our model. Other demographic, clinical and genotype variables, such as gender and ADAS11, can be viewed as the scalar response $Y$ in our model. See \cite{zhang2021high} for a detailed explanation and discussion of the ADNI DTI data.
We select gender, age, ADAS11 and ADAS13 as our scalar response candidates and test them separately. The estimated slope functions $\hat{\beta}(t)$ for each response are plotted in Figure \ref{fig3}.

The test statistics are $4.46\times10^{-5}$, $0.0619$, $0.0113$ and $0.2133$ with p-values $0.9946$, $0.8032$, $0.9157$ and $0.8838$, respectively. Therefore, at the significance leverl $0.05$, we do not reject the null hypothesis that FLM can be used to describe the data for the four responses.

\begin{figure*} [t!]
	\centering
	\subfloat[]{
		\resizebox{120pt}{!}{\includegraphics[width=0.4\textwidth]{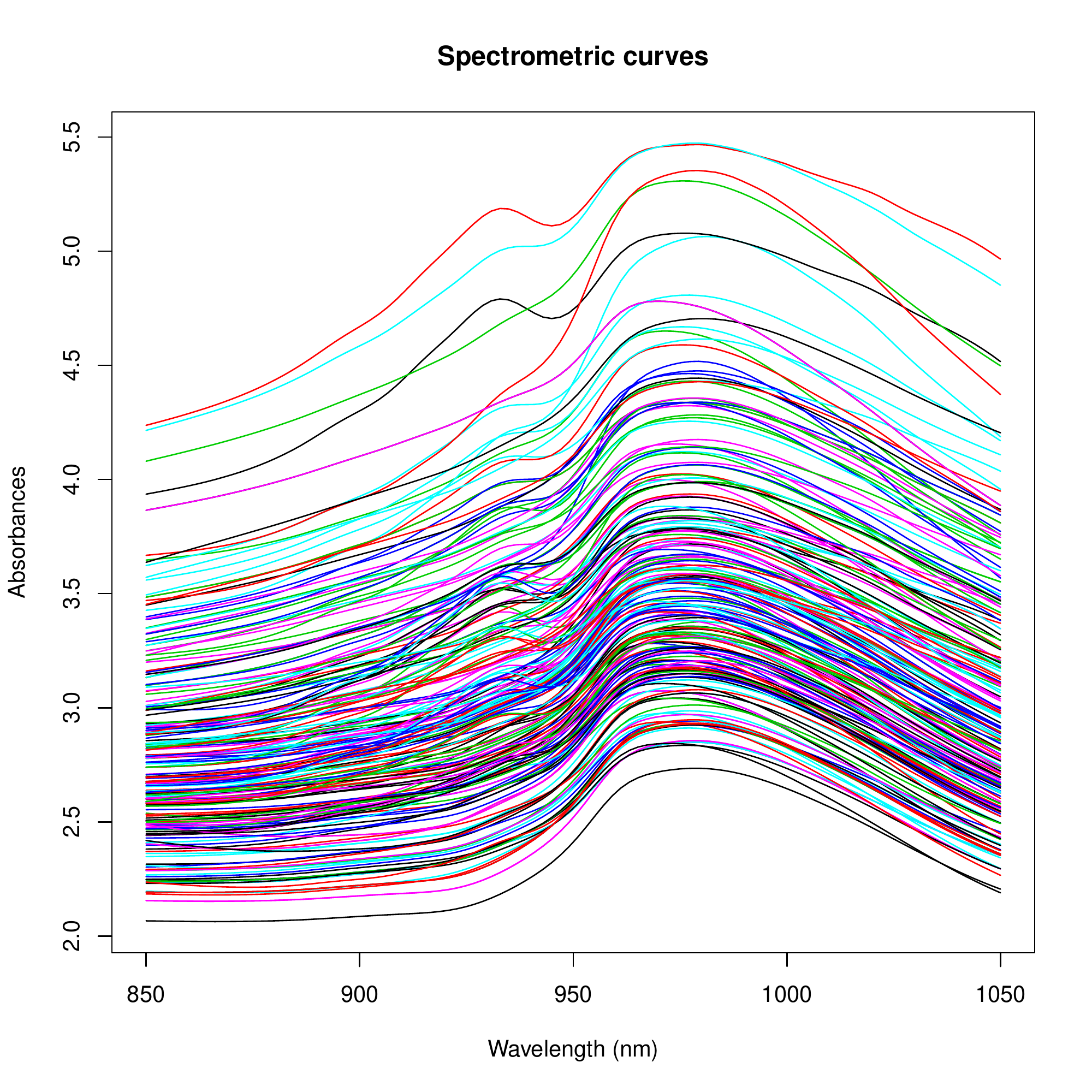}}}
	\subfloat[]{
		\resizebox{120pt}{!}{\includegraphics[width=0.4\textwidth]{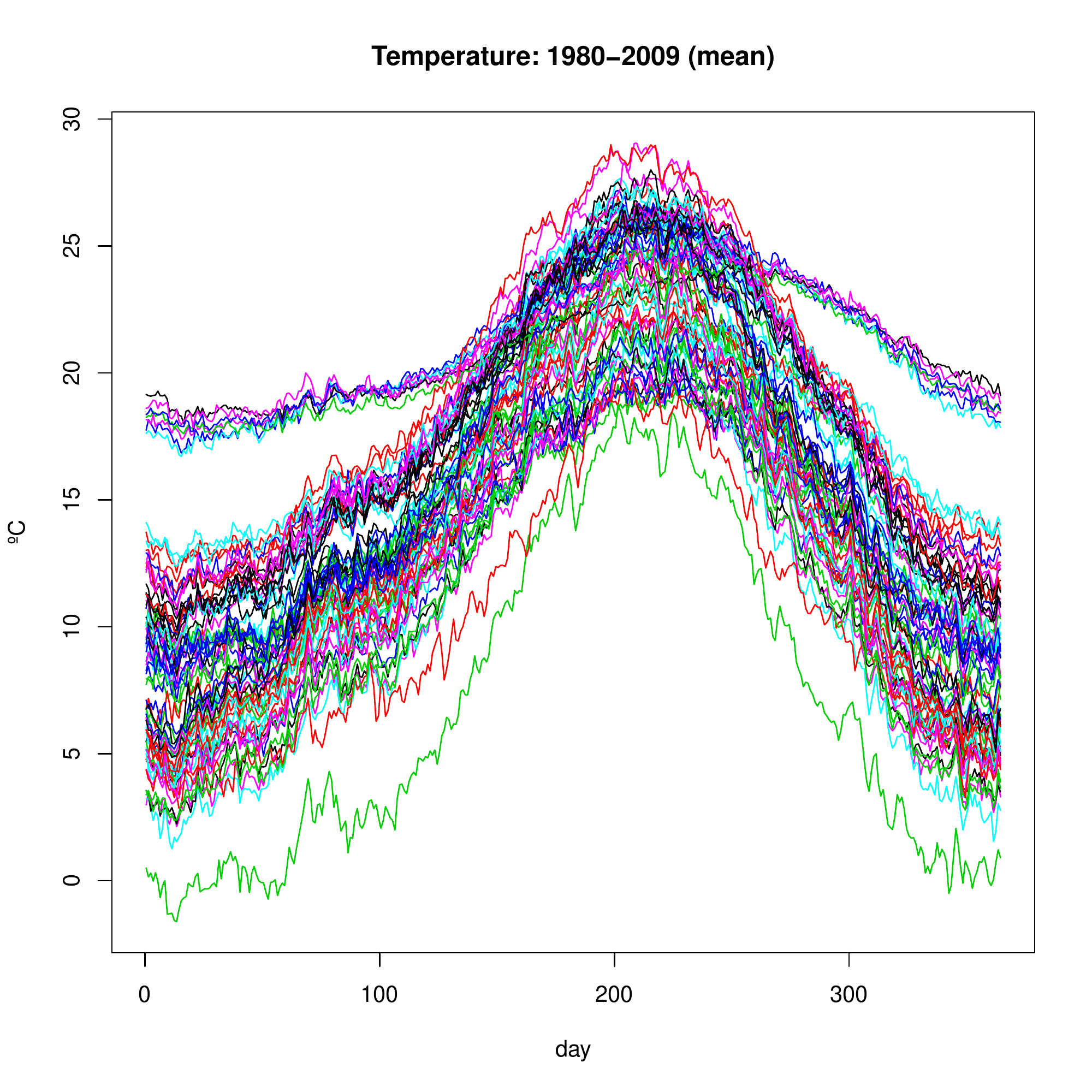}}}
	\subfloat[]{
		\resizebox{120pt}{!}{\includegraphics[width=0.4\textwidth]{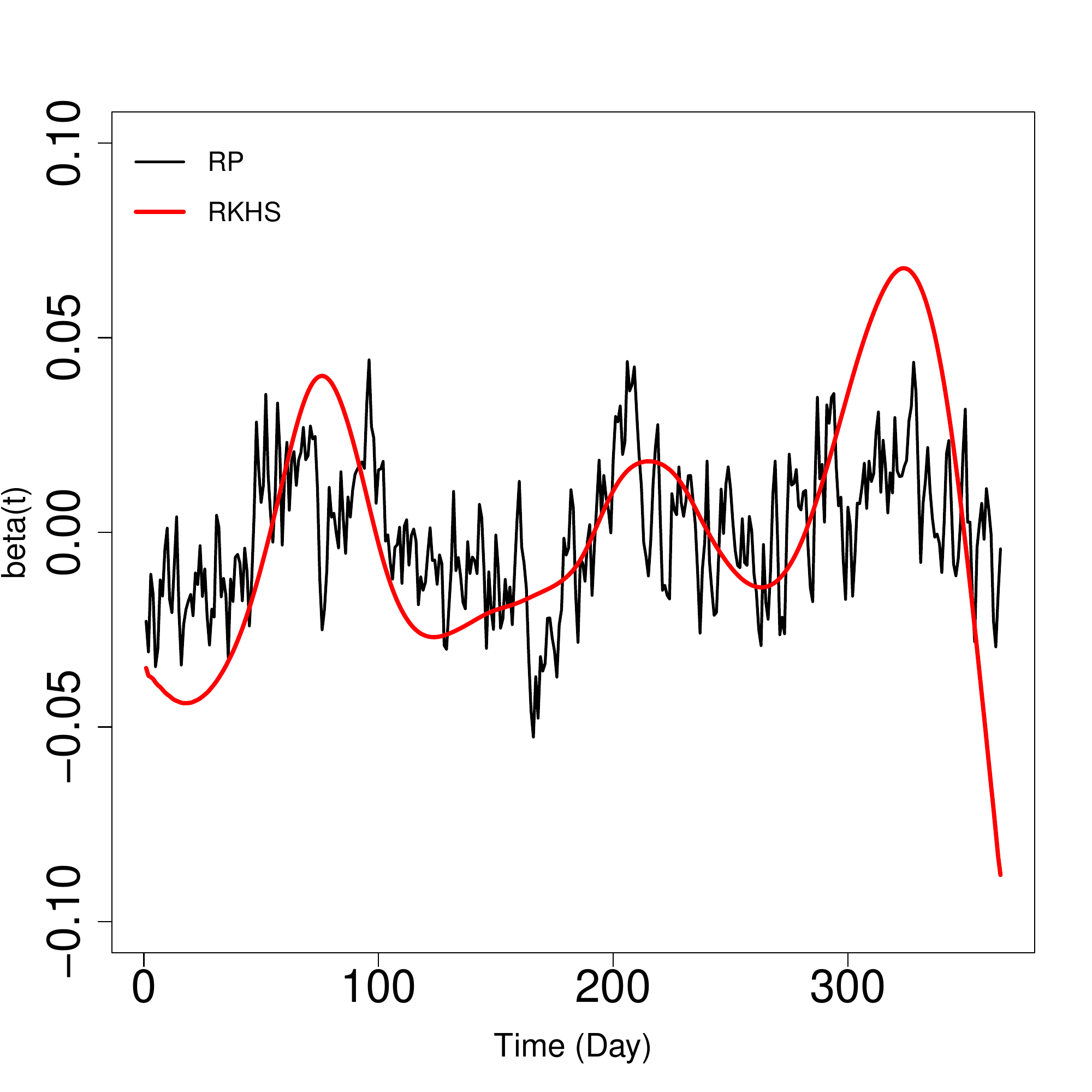}}}
\caption{(a) Tecator dataset with spectrometric curves; (b) AEMET temperatures for the 73 Spanish weather stations; and (c) The estimated
functional coefficient for the AEMET data set by random projection method (black) and RKHS method (red).}
\label{fig4}
\end{figure*}

The second and third examples we considered are the same as described in \cite{cuesta2019goodness}. Both datasets are provided in the R package \textbf{fda.usc}. Our proposed test returns the same conclusions. The second dataset is the classical Tecator data, a well known example in FDA for nonlinear regression.
The data record the absorbance of light at some particular wavelengths by 215 spectrometric curves collected from some chopped meat samples. The fat, water and protein contents of the meat samples are also included in the data set. We test whether these contents can be modeled by FLM using the spectrometric curves.
Before the test, six outliers are removed using the Fraiman and Muniz depth.
We test the adequacy of FLM for the data with the proposed $T_n$. The test statistics are $7.977\times10^3, 4.998\times10^4$ and $9.563\times10^3$ for fat, water and protein.
We reject the null hypothesis as the correspding $p$-values are all very close to $0$.
At the significance level $0.05$, there is strong statistical evidence against the FLM.
The third data we study is the Spanish weather stations data. The data contain yearly profiles of temperature from 73 weather stations of the AEMET (Spanish Meteorological Agency; Spanish acronym) network and other meteorological variables.
Our goal is to test whether the mean of the wind speed at each location can be described by FLM using the average yearly temperature curves.
We remove two outliers using the Fraiman and Muniz depth.
The test statistics is 1.149 and the p-value is 0.2856. Therefore, at the level $\alpha=0.05$, there is no evidence against the FLM.
The estimated slope functions for the two data sets are plotted in Figure \ref{fig4}.

\section{Discussion}

We propose an adaptive hybrid test for FLM. It is a powerful detector under the local and global alternatives with a tractable null distribution.
To achieve the adaptivity, we adopt an indicative dimension in an SDR subspace, which combines two simple but less powerful tests.
We develop an entirely data-driven method to determine the ridges in the ratio methods like TDRR to estimate the indicative dimension.
Besides, we consider the effect of the number of observations on a curve, which provides solid theoretical guarantees for real data applications.
Moreover, extensive numerical experiments demonstrate that our test can detect the local alternatives much better than existing methods and is competitive to its competitors under the global alternatives.
The proposed test is feasible to real data sets, and its validity is supported by the motivating example data set for nonlinear functional model.

We conclude the paper with several directions for future study.
First, although our method is proposed under the assumption of a functional linear model, it is possible to extend the results to other scenarios like a functional partial linear model or a nonlinear functional model.
For the other models, there have been works on parameter estimation. However, many of them lack the study of asymptotic properties. This results in challenges in deriving the asymptotic behaviours of the corresponding test statistic.
Second, the proposed test can achieve a faster convergence rate under the null if skipping the smoothing procedure for the discretely observed functional data. In the literature, the slope function can be estimated using the discretely observed functional data directly without smoothing them. In this case, the estimator may have a faster convergence rate, leading to a more powerful test under the alternatives.
A thorough investigation of estimation and inference problems based on discretely observed functional data is needed to achieve this.
Third, the data-driven method proposed for ridge selection is promising as existing approaches tend to underestimate the underlying dimension or rely on some manually determined ridges. An underestimated dimension results in the weaker ability to detect local and global alternatives. TDRR is proposed to solve this issue but still needs to select ridges. Based on our simulation experience, the dimension estimation is related to the data generating process due to the subspace construction. Therefore, to fully use the data structure and reduce the possible power loss caused by an underestimated dimension, we propose the data-driven method.

\bibliographystyle{unsrt}
\bibliography{FLM}




\end{document}